\documentclass[11pt]{article}

\usepackage{latexsym}
\usepackage{amsfonts}
\usepackage{amssymb}
\usepackage{amsmath}
\usepackage{epsf}

\newcommand{\bd}{\begin{displaymath}}
\newcommand{\ed}{\end{displaymath}}
\newcommand{\be}{\begin{equation}}
\newcommand{\ee}{\end{equation}}
\newcommand{\ba}{\begin{eqnarray}}
\newcommand{\ea}{\end{eqnarray}}

\begin{document}

\author{R Mart{\'\i}nez-Casado$^{1,2}$, J L Vega$^2$\footnote{Present
address: Biosystems Group, School of Computing, University of Leeds,
Leeds, LS2 9JT United Kingdom.}, A S Sanz$^2$ and S Miret-Art\'es$^2$ \\
$^1$Ruhr-Universit\"at Bochum, Lehrstuhl f\"ur
Physikalische Chemie I\\
D-44801 Bochum, Germany\\
$^2$Instituto de Matem\'aticas y F{\'\i}sica Fundamental\\
Consejo Superior de Investigaciones Cient{\'\i}ficas\\
Serrano 123, 28006 Madrid, Spain.}

\title{Quasi-elastic peak lineshapes in adsorbate diffusion on nearly
flat surfaces at low coverages:\\ the motional narrowing effect in Xe
on Pt(111)}

\date{}

\maketitle

\begin{abstract}
Quasi-elastic helium atom scattering measurements have provided
clear evidence for a two-dimensional free gas of Xe atoms on
Pt(111) at low coverages.
Increasing the friction due to the surface, a gradual change of the
shape of the quasi-elastic peak is predicted and analyzed for this
system in terms of the so-called motional narrowing effect.
The type of analysis presented here for the quasi-elastic peak should
be prior to any deconvolution procedure carried out in order to better
extract information from the process, e.g.\ diffusion coefficients and
jump distributions.
Moreover, this analysis also provides conditions for the free gas regime
different than those reported earlier.
\end{abstract}


\section{Introduction}
\label{sec1}

One of the main observables in adsorbate diffusion on metal
surfaces is the full width at half maximum (FWHM) of the
quasi-elastic (Q-)peak as a function of the parallel
momentum transfer, $\Delta {\bf K} = {\bf K}_f - {\bf K}_i$.
Different magnitudes of interest can be easily extracted from it,
e.g.\ diffusion coefficients and jump distributions.
In the 1950s van Hove \cite{vanHove} and Vineyard \cite{vineyard}
studied the broadening/narrowing of the Q-peak for several simple
models (ranging from the free particle to a lattice atom and the
liquid phase) in order to characterize the nature of the phase under
study in terms of this property (if possible).
However, no attention has been usually paid to the actual lineshape
of the Q-peak.
Is the shape of the Q-peak preserved when the surface friction and
the parallel momentum transfer of the particles probing the adsorbates
change?
This is a critical question; a precise knowledge of the natural
lineshape is crucial in the experimental deconvolution procedure for
a better description of the diffusion mechanisms underlying the
adsorbate dynamics.
Nowadays, the wealth of experimental data available in surface
science (obtained either by standard time-of-flight
techniques \cite{toennies1,toennies2} or by spin-echo
measurements \cite{jardine}) allows us to tackle this question
successfully, as we show in this work.
In particular, quasi-elastic helium atom scattering (QHAS) has
provided \cite{toennies2} the first experimental evidence for a fully
mobile two-dimensional gas of Xe atoms on Pt(111) at low coverage
($\theta = 0.017$), low incident helium atom energy ($E_i = 10.15$~meV)
and a surface temperature $T_s = 105$~K.

One of the theoretical approaches used to interpret the experimental
results from QHAS measurements is the classical Langevin formalism
\cite{vega,eli}, sometimes combined with standard molecular dynamics
techniques \cite{toennies2}.
Here we are going to use such an approach in order to analyze the
Xe-Pt(111) system at the experimental conditions considered by Ellis
{\it et al.}\ \cite{toennies2}, analyzing the Q-peak lineshape in the
light of the so-called motional narrowing effect
\cite{vega,vleck,kubo}.
As will be seen, this type of analysis should be prior to any
deconvolution procedure aimed to better extract information from
the process, such as diffusion coefficients or jump distributions.

The organization of this work is as follows.
In order to be self-contained, in section~\ref{sec2} we present the
theoretical approach that we have followed to obtain the analytical
formulae that will allow us to interpret our numerical results; the
numerical methodology is also presented in this section.
The results derived from our approach are shown and discussed in
section~\ref{sec3}.
Finally, the main conclusions are summarized in section~\ref{sec4}.


\section{Theoretical approach}
\label{sec2}

From QHAS experiments one obtains the differential reflection
coefficient which, in analogy to liquids \cite{vanHove,silvestri},
can be expressed as
\be
 \frac{d^2 {\mathcal R} (\Delta {\bf K}, \omega)}{d\Omega d\omega}
  = n_d {\mathcal F} S(\Delta {\bf K}, \omega) ,
 \label{eq:DRP}
\ee
and gives the probability that the He atoms scattered from the
diffusing collective reach a certain solid angle $\Omega$ with an
energy exchange $\hbar\omega =E_f - E_i$ and parallel (to the
surface) momentum transfer $\Delta {\bf K} = {\bf K}_f - {\bf K}_i$.
In the right hand side (r.h.s.) of equation~(\ref{eq:DRP}), $n_d$ is
the adparticle concentration; ${\mathcal F}$ is the atomic form factor,
which depends on the interaction potential between the probe atoms and
the adparticles; and $S(\Delta {\bf K},\omega)$ is the so-called
dynamic structure factor or scattering law, which is the observable
magnitude in this type of experiments and provides complete information
about the dynamics and structure of the ensemble of adparticles, and
therefore also about the surface diffusion process.
For example, information about long distance correlations is
obtained from $S(\Delta {\bf K},\omega)$ when considering small values
of $\Delta {\bf K}$, while long timescale correlation information is
available at small values of $\hbar \omega$.

The starting point of our approach consists of expressing the dynamic
structure factor as \cite{vanHove,vega,silvestri}
\be
 S(\Delta {\bf K},\omega) =
  \int e^{-i\omega t} \ \! I(\Delta{\bf K},t) dt ,
 \label{eq:DSF}
\ee
where
\be
 I(\Delta {\bf K},t) \equiv \langle e^{-i\Delta {\bf K} \cdot
   [{\bf R}(t) - {\bf R}(0)] } \rangle
  = \langle e^{-i \Delta K \!
   \int_0^t v_{\Delta {\bf K}} (t') dt'} \rangle
 \label{eq:IntSF}
\ee
is the intermediate scattering function, where the brackets denote
(ensemble) averaging over the adsorbates' trajectories, ${\bf R}(t)$.
In (\ref{eq:IntSF}), $v_{\Delta {\bf K}}$ is the adparticle velocity
projected onto the $\Delta {\bf K}$-direction (note that
$\Delta K \equiv \| \Delta {\bf K} \|$ is the length of
$\Delta {\bf K}$).

Both the dynamic structure factor and the intermediate scattering
function can be readily obtained from Langevin numerical simulations
as follows.
Within a low coverage regime (as happens in Ellis {\it et al.}\
experiment \cite{toennies2}, where $\theta = 0.017$),
adsorbate-adsorbate interactions can
be neglected, and diffusion is fully characterized by only studying
the dynamics of an isolated adsorbate on a metal surface.
This is the so-called single adsorbate approximation.
The adsorbate-substrate interaction includes the effects due to
the surface corrugation as well as those arising from the surface
thermal vibrational modes.
Considering a temperature dependent expansion, both effects can be
separated.
Thus, one the one hand, there is an adiabatic, periodic
adsorbate-substrate interaction potential $V$ which is temperature
independent, namely the zero-temperature potential (i.e.\
$T_s = 0$~K).
On the other hand, there is a coupling term accounting for the
vibrational effects induced by the temperature on the (surface)
lattice atoms that act on the adsorbate, which can be substituted
by a stochastic noise source \cite{gardiner}.
This allows to use a Langevin approach \cite{toennies1}, where the
force acting on the adsorbates is given by two contributions: (1) a
deterministic force $F = - \nabla V$, and (2) a stochastic force
$G(t)$.
As in standard Langevin molecular dynamics
simulations \cite{toennies1,toennies2}, the stochastic force
has features of a Gaussian white noise (the diffusion process
is considered as a Brownian-like motion \cite{gardiner}), i.e.
\ba
 \langle G(t) \rangle & = & 0 ,
 \label{cond1a} \\
 \langle G(t_1) G(t_2) \rangle & = &
  2m\gamma k_B T_s \delta(t_2 - t_1) ,
 \label{cond1b}
\ea
where $m$ is the adsorbate mass and $\gamma$ is the
adsorbate-substrate coupling strength or friction coefficient.

Taking into account the previous discussion, the motion of an
isolated adsorbate under the action of a bath consisting of the
temperature-dependent surface vibrations can be characterized
by the standard Langevin equation
\be
 m \ddot{{\bf R}}(t) =
  - m \gamma \dot{{\bf R}}(t) + F({\bf R}(t)) + {\bf G}(t) .
 \label{eq-lang1}
\ee
Here, ${\bf G}(t)$ is the two-dimensional (Gaussian white noise)
stochastic force, whose components satisfy the conditions given by
equations~(\ref{cond1a}) and (\ref{cond1b}) --moreover,
$\langle G_x(t_1) \ \! G_y(t_2) \rangle = 0$.
Note that this equation of motion is based on the Markovian
hypothesis: the correlations of the surface fluctuating force decay
very rapidly [see equation~(\ref{cond1b})].
On the other hand, also notice that $\gamma$ gives rise to a
characteristic timescale, the correlation time $\tau = 1/\gamma$,
which is related to the mean free path that an adsorbate can travel
without feeling much friction.

In order to interpret the Langevin numerical simulations, we
can express the intermediate scattering function, given by r.h.s.\ of
the second equality of equation~(\ref{eq:IntSF}), as a second order
cumulant expansion in $\Delta K$,
\be
 I(\Delta {\bf K}, t) \approx
  e^{- \Delta K^2 \int_0^t (t - t') \mathcal{C}_{\Delta {\bf K}}(t') dt'} ,
 \label{eq:IntSF2}
\ee
where $\mathcal{C}_{\Delta {\bf K}} (t) \equiv
\langle v_{\Delta {\bf K}}(0) \ \! v_{\Delta {\bf K}}(t) \rangle$
is the autocorrelation function of the velocity projected onto
the direction of the parallel momentum transfer.
This is the so-called Gaussian approximation \cite{mcquarrie},
which is exact when the velocity correlations at more than two
different times are negligible.
Despite its limitations, it provides much insight into the dynamical
process by allowing an almost analytical treatment of the problem.
Note that it allows to replace the average acting over the exponential
function by an average acting over its argument, thus simplifying the
analytical derivation in some simple but relevant cases.

For an almost flat surface ($V \approx 0$), any direction is equivalent
and therefore the dimensionality of the numerical Langevin simulation
reduces to one.
The corresponding numerical velocity autocorrelation function then
follows the exponential behavior
\be
 \mathcal{C}(t) = \langle v_0^2 \rangle \ \! e^{- \gamma t} ,
 \label{corrGM2}
\ee
where $\langle v_0^2 \rangle = k_B T_s/m$ (the square root of this
magnitude gives the thermal velocity in one dimension).
Introducing (\ref{corrGM2}) into equation~(\ref{eq:IntSF2}), we
obtain \cite{kubo}
\be
 I(\Delta {\bf K}, t) = \exp \left[- \chi^2
   \left( e^{- \gamma t} + \gamma t - 1 \right) \right] ,
 \label{eq:IntSGM}
\ee
where $\chi$ is the shape parameter, defined as
\be
 \chi \equiv \sqrt{\langle v_0^2 \rangle}
  \ \! \Delta K  / \gamma = \bar{l} \ \! \Delta K .
 \label{chi}
\ee
From this relation the mean free path results
$\bar{l} \equiv \tau \sqrt{ \langle v^2 \rangle}$, and the diffusion
coefficient is $D \equiv \tau \langle v^2 \rangle$, which related to
the friction (Einstein's relation).
It can be easily shown \cite{vega} that the dynamic structure factor
derived from the Fourier transform of equation~(\ref{eq:IntSGM}) has a
different shape depending on the value of $\chi$.
This can be seen from the exact (analytical) Fourier transform of
equation~(\ref{eq:IntSGM}), which renders the following functional form
for the dynamic structure factor ruling the shape of the Q-peak
\ba
 S(\Delta {\bf K}, \omega) & = & \frac{\tau e^{\chi^2}}{\pi} \ \!
  \chi^{-2\chi^2} {\rm Re} \left\{ \chi^{-2i\omega \tau}
  \left[ \tilde{\Gamma} (\chi^2 + i\omega \tau)
   - \tilde{\Gamma} (\chi^2 + i\omega \tau, \chi^2) \right] \right\}
 \nonumber \\
 & = & \frac{e^{\chi^2}}{2 \pi}
  \sum_{n=0}^\infty \frac{(-1)^n \chi^{2n}}{n!}
  \frac{2(\chi^2 + n)/\tau}{\omega^2 + [(\chi^2 + n)/\tau]^2} \ \! ,
 \label{exact}
\ea
expressed in terms of the Gamma and incomplete Gamma functions
(denoted by $\tilde{\Gamma}$ in the r.h.s.\ of the first line of
equation~(\ref{exact}) to avoid any confusion with the FWHM,
$\Gamma$), respectively.
According to this expression, as $\chi$ decreases (or, equivalently,
as $\Delta K$ decreases, for $T_s$ and $\gamma$ fixed) the shape
of the dynamic structure factor goes from a purely Gaussian function
to a Lorentzian one, and its width gets narrower and narrower (see
appendix).
This is the so-called motional narrowing effect, well-known in the
theory of nuclear magnetic resonance lineshapes \cite{vega,vleck,kubo}.
The shape parameter goes from zero to infinity.
At high values of $\chi$, the Q-peak approaches a Gaussian shape,
which will be well reproduced only by few terms ($n$ not too small) in
the sum given by equation~(\ref{exact}).
For values of $\chi \ll 1$, the Q-peak approaches a Lorentzian shape,
with the term corresponding to $n=0$ usually being the dominant one
in equation~(\ref{exact}).
This variation between a Gaussian and a Lorentzian shape is in clear
correspondence with having either a ballistic (free particle) or a
diffusive regime, respectively.
Thus, a simple manner of expressing the FWHM of the Q-peak, $\Gamma$,
in terms of $\chi$ is
\be
 \Gamma = 2 \mu \gamma \chi^2
  + 2 \sqrt{2\ln 2} \ \! (1 - \mu) \ \! \gamma \chi \ \! ,
 \label{FWHM}
\ee
with $\mu$ a free parameter.
For $\mu = 0$, one obtains the FWHM of the Gaussian lineshape, while
the width corresponding to the Lorentzian lineshape is obtained for
$\mu = 1$ and when only the $n=0$ contribution of the infinite
sum in equation~(\ref{exact}) is considered (see appendix).
That is, as the importance of the diffusive regime increases, one
passes from a linear dependence on $\chi$ to a quadratic one.
The same behavior is observed in $\Gamma$ when it is written as a
function of $\Delta K$ because of the linear relationship between
this magnitude and $\chi$ (see equation~(\ref{chi})).

\begin{figure}
 \begin{center}
 \epsfxsize=3in {\epsfbox{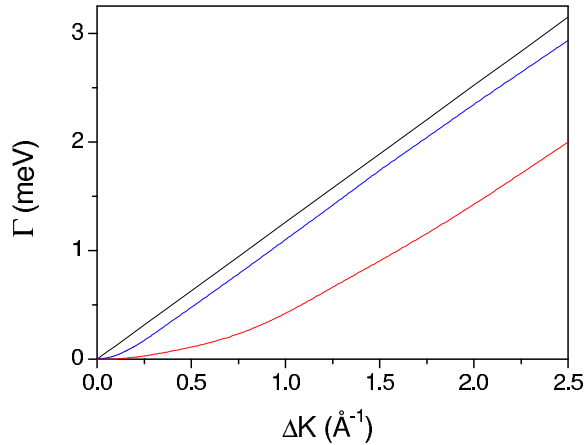}}
 \caption{\label{fig1}
  Full width at half maximum, $\Gamma$, of the quasi-elastic peak as
  a function of the parallel momentum transfer, $\Delta K$, along the
  direction (100) of the Pt(111) surface for three different values of
  the friction coefficient: $\gamma = 0$ (black/thick black line),
  $\gamma = 0.25$~ps$^{-1}$ (blue/thin black line), and
  $\gamma = 2.0$~ps$^{-1}$ (red/thin grey line).}
 \end{center}
\end{figure}


\section{Results and discussion}
\label{sec3}

\begin{figure}
 \begin{center}
 \epsfxsize=3in {\epsfbox{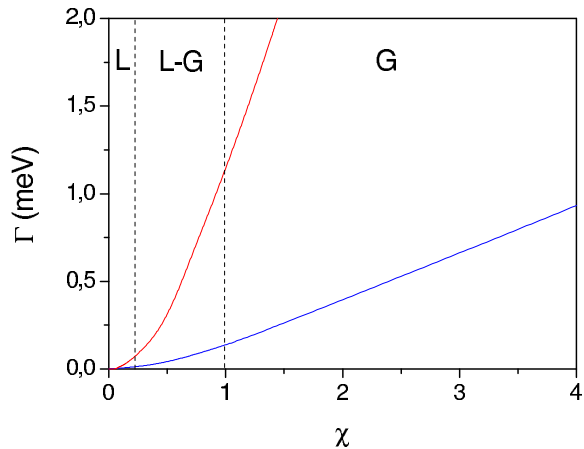}}
 \caption{\label{fig2}
  Full width at half maximum, $\Gamma$, of the quasi-elastic peak
  as a function of the shape parameter for two different values of
  the friction coefficient: $\gamma = 0.25$~ps$^{-1}$ (blue/thin black
  line) and $\gamma = 2.0$~ps$^{-1}$ (red/thin grey line).
  Vertical lines give the borders where a Lorentzian (L), a mixed
  Lorentzian-Gaussian (L-G), or a Gaussian (G) shape are observed.}
 \end{center}
\end{figure}

In figure~\ref{fig1}, $\Gamma$ is plotted as a function of the parallel
momentum transfer $\Delta K$ along the direction (100) for the
Xe-Pt(111) system at $T_s = 105$~K and three different values of the
friction coefficient (0, 0.25, and 2~ps$^{-1}$), assuming the
corrugation of the Pt(111) surface negligible.
These results are in excellent agreement with those obtained by Ellis
{\it et al.} \cite{toennies2} by means of Langevin molecular dynamics
simulations, i.e.\ considering interacting adsorbates, though the
interaction among them is relatively weak because the coverage used
in the experiment was relatively low ($\theta = 0.017$).
The experimental observation fitted perfectly the black solid line
(see figure~\ref{fig1}), this confirming what they claimed as a
two-dimensional free gas (zero friction).
In figure~\ref{fig1} it is also apparent a clear smooth transition in
$\Gamma$, from a quadratic to a linear dependence on $\Delta K$, in
accordance with equation~(\ref{FWHM}); the motional narrowing effect is
clearly observed as the friction coefficient is increased for a given
value of $\Delta K$.
The same behavior is also seen in figure~\ref{fig2}, where $\Gamma$ has
been plotted as a function of the shape parameter.
This figure also provides an important additional information: the
gradual change of the shape of the Q-peak as a function of $\chi$.
Very often, in order to extract information about the diffusion
mechanism, a Lorentzian shape is assumed to deconvolute the
experimental results.
As clearly seen in figure~\ref{fig2}, the pure Lorentzian shape is supposed
to be good in a very narrow range of $\chi$ (or $\Delta K$) values.
The three regions dividing this figure have been chosen according
to the fittings of the numerical results obtained from the Langevin
simulation to the analytical ones, given by equation~(\ref{exact}).
Though the two perpendicular border lines denoting the different
shapes are somewhat arbitrary, $\chi$ values greater than one will
give rise to a very strong Gaussian behavior.
Therefore, this kind of representation has the advantage that
experimentalists can use it as a guide in the deconvolution
procedure usually carried out within this context.

The smallest experimental accessed value of the parallel momentum
transfer, $\Delta K = 0.21$~\AA$^{-1}$, corresponds to a distance of
about 30~\AA, which should be regarded as an upper limit to the mean
free path of Xe atoms (or free gas regime) after reference
\cite{toennies2}.
Moreover, the experimental data for this system indicated that the Xe
atoms run freely along the surface, with upper limits of 0.25~ps$^{-1}$
for the friction and 9~meV for the diffusion barrier (at least less
than the thermal energy).
According to our theoretical study, the estimation of the mean free
path should be carried out in a different way.
The shape parameter and the friction can be easily obtained from a
proper fit of the Q-peak to equation~(\ref{exact}), and then the mean
free path is finally computed using equation~(\ref{chi}).
For example, for $\Delta K = 0.21$~\AA$^{-1}$ and
$\gamma = 0.25$~ps$^{-1}$, one obtains that $\bar{l} = 3.2$~\AA\ and
that the corresponding lineshape of the Q-peak displays a mixture
of Lorentzian and Gaussian shapes.

\begin{figure}
 \begin{center}
 \epsfxsize=3in {\epsfbox{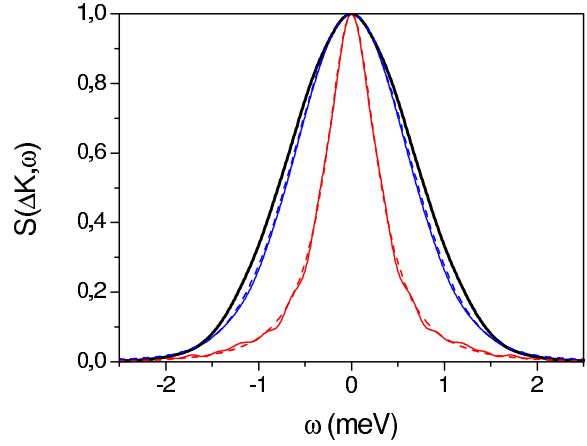}}
 \caption{\label{fig3}
  Dynamic structure factor, $S(\Delta {\bf K}, \omega)$, as a function
  of $\omega$ for $\Delta K = 1.25$~\AA$^{-1}$ and three different
  va\-lues of the friction coefficient: $\gamma = 0$ (black/thick black
  line), $\gamma = 0.25$~ps$^{-1}$ ($\chi = 4.1$) (blue/thin black line),
  and $\gamma = 2.0$~ps$^{-1}$ ($\chi = 0.5$) (red/thin grey line).
  Dashed lines are fittings to the numerical lineshapes according to
  equation~(\ref{exact}).}
 \end{center}
\end{figure}

\begin{figure}
 \begin{center}
 \epsfxsize=3in {\epsfbox{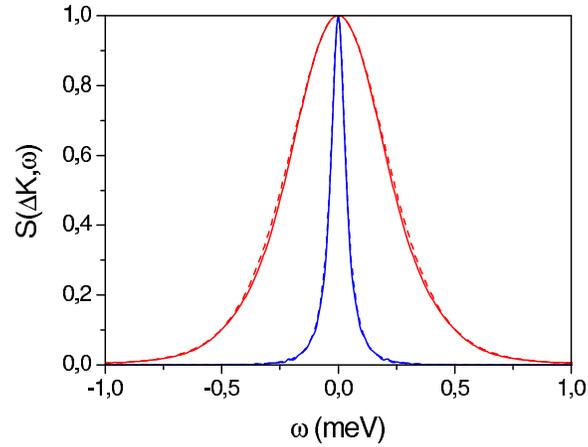}}
 \caption{\label{fig4}
  Dynamic structure factor, $S(\Delta {\bf K}, \omega)$, as a function
  of $\omega$ for $\gamma = 0.25$~ps$^{-1}$ and two different values
  of $\Delta K$: $\Delta K = 0.15$~\AA$^{-1}$ ($\chi = 0.5$) (blue/thin
  black line) and $\Delta K = 0.5$~\AA$^{-1}$ ($\chi = 1.6$) (red/thin
  grey line).
  Dashed lines are fittings to the numerical lineshapes according to
  equation~(\ref{exact}).}
 \end{center}
\end{figure}

The motional narrowing effect is also clearly seen in figures~\ref{fig3}
and \ref{fig4}, where numerical quasi-elastic lineshapes corresponding
to different values of $\chi$ (or $\Delta K$) are shown. Dashed lines
are fittings to the numerical lineshapes according to equation~(\ref{exact}).
In particular, in figure~\ref{fig3}, the thick black curve corresponds
to a zero friction case, for which the numerical (and analytical)
lineshape is a pure Gaussian function.
The blue curve is obtained from a Langevin calculation with a friction
$\gamma = 0.25$~ps$^{-1}$, and the best fitting to equation~(\ref{exact})
is reached with $n_{max}=58$, approaching a Gaussian shape.
Conversely, the red curve displays a perfect Lorentzian behavior
($n=0$, with a very slight correction of the lowest part of the
Lorentzian that comes from $n=1$).
Analogously, in figure~\ref{fig4}, the numerical blue and red
curves fit to a perfect Lorentzian ($n=0$, with an also
slight correction arising from $n=1$) and a mixed Lorentian-Gaussian
function ($n_{max}=10$), respectively.

\section{Conclusions}
\label{sec4}

As seen in this work, the results obtained from the numerical Langevin
simulations carried out by Ellis {\it et al.}\ \cite{toennies2} fit
perfectly to the analytical formalism based on the
Gaussian approximation of the intermediate scattering function.
This allows to study the lineshape of the Q-peak in a low coverage
regime and for very low corrugated surfaces, as it is the case for
the Pt(111) surface.
Accordingly, we have shown that a gradual change of the shape of the
Q-peak is expected when increasing the friction and $\Delta K$, which
can be understood in the light of the so-called motional narrowing
effect.
This type of analysis of the Q-peak should be therefore prior to
any deconvolution procedure carried out in order to better extract
information from the process (e.g.\ diffusion coefficients and
jump distributions).
Moreover, taking into account these numerical simulations, conditions
for the corresponding free gas regime quite different to those reported
earlier have been found.


\vspace{1cm}
\noindent {\bf \large Acknowledgments}
\vspace{.5cm}

\noindent
This work was supported in part by DGCYT (Spain), Project No.\
FIS2004-02461.
RM-C would like to thank the University of Bochum for support
from the Deutsche Forschungsgemeinschaft, SFB 558, for a predoctoral
contract.
JLV and AS Sanz would like to thank the Ministry of Education
and Science (Spain) for a predoctoral grant and a ``Juan de la Cierva''
Contract, respectively.


\appendix
\setcounter{section}{1}

\section*{Appendix}
 \label{append}

In this appendix we analyze the two extreme cases of the dynamic
structure factor lineshape (Lorentzian and Gaussian) as a function
of the shape parameter, starting from equation~(\ref{exact}).

For $\chi \ll 1$, the dominant term in the second equality of
equation~(\ref{exact}) is that corresponding to $n=0$.
Therefore, one obtains straightforwardly that
\be
 S(\Delta {\bf K}, \omega) \approx \frac{1}{\gamma\chi^2}
   \frac{1}{1 + (\omega/\gamma\chi^2)^2} ,
 \label{lineshapes1}
\ee
which is a Lorentzian function with FWHM $\Gamma = 2\gamma\chi^2 =
2D\Delta K^2$ (i.e.\ $\mu = 1$ in equation~(\ref{FWHM})).

On the other hand, for large $\chi$, it is convenient to start with
the first equality of equation~(\ref{exact}), which is written in terms
of the incomplete Gamma and Gamma functions \cite{grad}, whose
asymptotic behaviors are
\ba
 \tilde{\Gamma} (\alpha,\beta) & \approx &
  \beta^{\alpha - 1} e^{-\beta} ,
 \label{igamma1} \\
 \tilde{\Gamma} (\alpha) & \approx &
  \sqrt{2\pi} \ \! \alpha^{\alpha - 1/2} e^{-\alpha} ,
 \label{gamma1}
\ea
respectively, with $\alpha = \chi^2 + i \omega/ \gamma$ and
$\beta = \chi^2$.
For the sake of simplicity, it is better to express $\alpha$ is its
polar form, $\alpha = \rho \ \! e^{i\delta}$, with
\be
 \rho = \sqrt{\chi^4 + \left(\frac{\omega}{\gamma}\right)^2} ,
 \quad \delta = (\tan)^{-1} \left( \frac{\omega/\gamma}{\chi^2} \right)
  \approx \frac{\omega}{\gamma\chi^2}
\ee
(the approximation in $\delta$ arises after assuming large $\chi$).
In doing so, equations~(\ref{igamma1}) and (\ref{gamma1}) become
\be
 \tilde{\Gamma} (\chi^2 + i\omega/\gamma, \chi^2) \approx
   \chi^{2(\chi^2 + i\omega/\gamma)} e^{-\chi^2} \chi^{-2}
 \label{igamma2}
\ee
and
\ba
 \tilde{\Gamma} (\chi^2 + i\omega/\gamma) = \sqrt{2\pi}
  \left[ \left[ \chi^4 + \left(\frac{\omega}{\gamma}\right)^2
   \right]^{1/2} \right]^{\chi^2} \nonumber \\
   \times e^{-\omega^2/\gamma^2\chi^2}
  e^{i\{\omega/\gamma + (\omega/2\gamma)
   \ln [\chi^4 + (\omega/\gamma)^2]\} } \nonumber \\
  \times
   \left[ \chi^4 + \left(\frac{\omega}{\gamma}\right)^2 \right]^{-1/4}
  e^{-i\omega/2\gamma\chi^2} e^{- (\chi^2 + i\omega/\gamma)} ,
 \label{gamma2}
\ea
respectively.
After some straightforward algebraic manipulations, equation~(\ref{gamma2})
can be rewritten as
\be
 \tilde{\Gamma} (\chi^2 + i\omega/\gamma) \approx
 \sqrt{2\pi} \ \! \chi^{2(\chi^2 + i\omega/\gamma)} e^{-\chi^2}
  \chi^{-1} e^{-\omega^2/2\gamma^2\chi^2} .
 \label{gamma5}
\ee
Substituting (\ref{igamma2}) and (\ref{gamma5}) in equation~(\ref{exact})
we finally obtain
\be
 S(\Delta {\bf K}, \omega) \approx \frac{1}{\gamma}
  \left[ \sqrt{2\pi} \chi^{-1} e^{-\omega^2/2\gamma^2\chi^2}
   - \chi^{-2} \right] ,
 \label{exact2}
\ee
which in the limit of large $\chi$ reads as
\be
 S(\Delta {\bf K}, \omega) \approx \frac{\sqrt{2\pi}}{\gamma\chi}
  \ \! e^{-\omega^2/2\gamma^2\chi^2} .
 \label{exact3}
\ee
This is a Gaussian function with $\Gamma = 2 \sqrt{2\ln 2} \ \!
\gamma \chi \propto \Delta K$ (i.e.\ $n \to \infty$ in
equation~(\ref{exact}) and $\mu = 0$ in equation~(\ref{FWHM})).


\end{document}